\documentstyle[prd,tighten,aps]{revtex}
\begin{document}
\draft

\twocolumn[\hsize\textwidth\columnwidth\hsize\csname
@twocolumnfalse\endcsname
\title{\bf Cosmological Models from Quintessence}

\author{Pedro F. Gonz\'{a}lez-D\'{\i}az}
\address{Isaac Newton Institute, 20 Clarkson Road, Cambridge, CB3 0EH, UK\\
and\\
Centro de F\'{\i}sica ``Miguel Catal\'{a}n'', Instituto de
Matem\'{a}ticas y F\'{\i}sica Fundamental,\\ Consejo Superior de
Investigaciones Cient\'{\i}ficas, Serrano 121, 28006 Madrid (SPAIN)}
\date{April 5, 2000}

\maketitle

\begin{abstract}
A generalized quintessence model is presented which corresponds
to a richer vacuum structure that, besides a time-dependent,
slowly varying scalar field, contains a varying cosmological
term. From first principles we determine a number of
scalar-field potentials that satisfy the constraints imposed by
the field equations and conservation laws, both in the
conventional and generalized quintessence models. Besides
inverse-power law solutions, these potentials are given in terms
of hyperbolic functions or the twelve Jacobian elliptic
functions, and are all related to the luminosity distance by
means of a integral equation. Integration of this equation for
the different solutions leads to a large family of cosmological
models characterized by luminosity distance-redshift relations.
Out of such models, only four appear to be able to predict a
required accelerating universe conforming to observations on
supernova Ia, at large or moderate redshifts.
\end{abstract}

\vskip2pc]

\renewcommand{\theequation}{\arabic{section}.\arabic{equation}}

\section{\bf Allowed potentials for quintessence}
\setcounter{equation}{0}

Quintessence has been recently invoked [1] as an advantageous
alternative to the cosmological constant in order to explain the
apparent accelerating expansion of the universe which has
stirred cosmologists after observations and measurements of
distant supernovae [2-4]. The existence of a quintessential
field has been related to supersymmetric models [5], the problem
of fine-tuning of the cosmological constant [6], or supergravity
models [7]. The bare standard cosmological model (BSCM), without
any constant cosmological term or vacuum fields, predicts the
existence of an expanding universe which can be closed, open or
flat, but always decelerating. However, if a positive
cosmological constant is added to the field equations, then the
expansion of the universe may become accelerating. Actually, as
early as 1975, Gunn and Tinsley, while discussing observations
on the Hubble diagram and constraints on the matter density of
the universe and ages of galaxies, found [8] a series of
allowable, if not compelling, cosmological models with nonzero,
positive cosmological constant, which were accelerating. Recent
observations [2,3] on distant supernovas have resurrected the
spirit of these early conclusions and led to the strong suspect
that, in spite of the fact that the BSCM gives satisfactory
explanations to many other observations, it probably is
incomplete or even incorrect [9].

When the cosmological constant, $\Lambda$, is interpreted as the
energy density of vacuum for an equation of state $p=-\rho$, if
it is positive, then most inflationary models are suitably
pinpointed. However, it is largely known that $\Lambda$ is not
free of fundamental problems [10]. Actually, the so-called
cosmological constant problem is one of the most challenging
questions in fundamental physics, as it is very hard to envisage
any consistent mechanism that dynamically explains how the
vacuum energy density can be lowered from its most natural value
at around the Planck scale down to its observationally allowed
value, $\epsilon \leq 10^{-47}$ GeV$^{4}$. Although quintessence
models do not solve this problem, they may improve the related
fine-tuning problem in the sense that [6] they can explain a
tiny value for the vacuum energy density with a scale comparable
with the scales of high energy physics. Besides, these models
give rise to an accelerating universe by using a vacuum
dynamically adjustable, time-dependent scalar field that is
spatially (in-)homogeneous and evolves slowly enough so that the
kinetic term of the energy density is always smaller than the
potential energy term. It is worth noticing that this is not
necessarily required in tracker models of quintessence (see e.g
Ref. [15]). Indeed, in the case of an overshoot [15] the kinetic
energy dominates at high redshift. If we disregard such tracker
models, the resulting negative pressure will then correspond to
an equation of state [1] $p=\omega\rho$ where the free-parameter
$\omega$ can take on any values $0\geq\omega>-1$. Thus, the
cosmological constant will correspond to the extreme case
$\omega=-1$.

Recently, however, di Prieto and Demaret have shown [11] that,
if we restrict ourselves to a constant equation of state none of
the vacuum scalar-field potentials, $V(\phi)$, currently used in
quintessence models, such as the exponential [12], cosine [13]
and inverse power-law [12] potentials can satisfy the constraint
on $V(\phi)$ implied by field equations and conservation laws,
i.e. [11]
\[\frac{V'}{V_0 '}=\]
\begin{equation}
\pm\left[\Omega_{\phi}\left(\frac{V}{V_0}\right)^2
+\Omega_M\left(\frac{V}{V_0}\right)^{\frac{\omega+2}{\omega+1}}
+\Omega_k\left(\frac{V}{V_0}\right)^{\frac{3\omega+5}{3(\omega+1)}}
\right]^{\frac{1}{2}} ,
\end{equation}
where $V_0$ and $V_0'$ are the current values of the
scalar-field potential, $V(\phi)$, and its derivative with
respect to the field, $V'=dV(\phi)/d\phi$, respectively, and the
$\Omega_i$'s (with $i=\phi,M,k$) are the dimensionless density
parameters for the scalar field, ordinary matter and topological
curvature. Apart from a solution for any $\omega$ in the flat
case, Pietro and Demaret were nonetheless able to find [11] some
solutions to constraint (1.1) for particular values of the
parameter $\omega$. Thus, for $\omega=-1/3$, they obtained
\begin{equation}
V(\phi)= V_0\left\{\frac{\sqrt{2}}{4\epsilon_0}
\sinh\left[\pm\epsilon_0(\phi-\phi_0)+\nu_0\right]\right\}^{-4},
\end{equation}
where
\[\epsilon_0=\frac{1}{2}\sqrt{\frac{\Omega_{\phi}+\Omega_k}{2\Omega_M}}
,\;\; \nu_0=\arcsin\left(2\sqrt{2}\epsilon_0\right) ,\] and, for
$\omega=-2/3$,
\[V(\phi)=V_0
\left\{\frac{\Omega_M}{\Omega_{\phi}}\sinh\left[\pm(\phi-\phi_0)
+\delta_0\right]\right.\]
\begin{equation}
\left. +\frac{\Omega_k}{4\Omega_{\phi}}
\left(\frac{\Omega_k}{\Omega_M}e^{\mp(\phi-\phi_0)-\sigma_0}
-2\right) \right\}^{-\frac{1}{2}} ,
\end{equation}
where
\[e^{\sigma_0}=
\frac{2\Omega_{\phi}+2\sqrt{\Omega_{\phi}}+\Omega_k}{\Omega_M}\]
\[\delta_0=\sigma_0+\frac{1}{2}\ln\left(\frac{\Omega_M}{4\Omega_{\phi}}\right)
.\]

We furthermore note that, besides solutions (1.2) and (1.3),
there are a whole family of scalar potentials $V(\phi)$ defined
in terms of the Jacobian elliptic functions [11], $J_e$, which
satisfy the constraint (1.1) for $\omega=-1/6$. Such solution
can be generally written as
\begin{equation}
V(\phi)=V_0\left\{J_e\left[\alpha_0(\phi-\phi_0),m\right]\right\}^{-10}
,
\end{equation}
in which $\alpha_0=\alpha_0(\Omega_i)$ is a given constant whose
form depends on the particular elliptic function being
considered, and $m=m(\Omega_i)\leq 1$ is the characteristic
parameter (modulus) [14] of the corresponding elliptic function.
For example, taking for $J_e$ the function {\rm cn}, we have
\[\alpha_0=
\sqrt{\frac{7\Omega_k(\Omega_{\phi}+2\Omega_k)}{200\Omega_{\phi}(\Omega_k+\Omega_{\phi})}},\]
\[m=\frac{\Omega_k}{\Omega_{\phi}+2\Omega_k},\]
or for the function {\rm sd}
\[\alpha_0=\sqrt{\frac{7\Omega_M}{200\Omega_{\phi}}},\]
\[m=\frac{1}{2}\left(1+\frac{2\Omega_{\phi}^2}{7}\right),\]
and similar, but distinct expressions of $\alpha_0$ and $m$ for
the remaining 10 Jacobian elliptic functions.

It appears clearly of interest to investigate whether the new
potentials (1.2)-(1.4) are actually able to predict accelerating
cosmological models which can match recently obtained data from
observations on distant supernova Ia, discussing their physical
relevance as well. It is the aim of this paper to carry out such
an investigation, incorporating other possible new solutions
from a generalized quintessence model which simultaneously
accommodates both a vacuum scalar field $\phi$ and a varying
cosmological term $\Lambda$. In this paper, we shall restrict
ourselves to equations of state with a constant $\omega$, both
in conventional and generalized quintessence models,
disregarding tracker models [15], where time varying equations
of state are invoked and a general inverse-power law potential
for the quintessence field is assumed.

\section{\bf Generalized quintessence model}
\setcounter{equation}{0}

The field equations corresponding to a
Friedmann-Robertson-Walker spacetime with ordinary matter which
is not coupled to a homogeneous (quintessence) scalar field
$\phi$, to which we add a varying cosmological term $\Lambda$
can be written as
\begin{equation}
\frac{\dot{R}^2}{R^2}+\frac{k}{R^2}=
\frac{1}{3}\kappa^2\left(\rho_{\phi}+\rho_M+\rho_{\Lambda}\right)
\end{equation}
\begin{equation}
2\frac{\ddot{R}}{R}+\frac{\dot{R}^2}{R^2}+\frac{k}{R^2}=
\kappa^2\left(p_{\Lambda}-p_{\phi}\right)
\end{equation}
\begin{equation}
\ddot{\phi}+3\dot{\phi}\frac{\dot{R}}{R}=-V' ,
\end{equation}
where the overhead dot means time derivative, $'\equiv d/d\phi$,
$\kappa^2=8\pi G_N$, $k$ is the topological curvature and we
have defined the scalar field such that
\begin{equation}
\kappa^2\rho_{\phi}=\frac{1}{2}\dot{\phi}^2+V(\phi)
\end{equation}
\begin{equation}
\kappa^2 p_{\phi}=\frac{1}{2}\dot{\phi}^2-V(\phi) .
\end{equation}
As usual, the ordinary matter is assumed to obey the equation of
state for an ordinary fluid, $p_M=0$. As pointed out before, the
scalar quintessence field will be assumed to behave like a
perfect fluid with equation of state [1]
\begin{equation}
p_{\phi}=\omega\rho_{\phi} ,\;\;\; -1<\omega\leq 0 .
\end{equation}
The generalization implied by the quintessence field with
respect to the case of a pure cosmological constant can be
manisfested by noting that the particular value of the constant
parameter $\omega=-1$ corresponds to the cosmological constant
case when the field $\phi$ becomes a constant as well [16].

The conservation laws that the involved fields are here assumed
to satisfy are as follows. First of all, we note that, since
there is no interaction between the scalar field and the other
fields involved, we can take all these laws as separable from
each other. For ordinary matter, $M$, and scalar field, $\phi$,
we should then have for all values of $\omega$, except
$\omega=-1$ [11]
\begin{equation}
\rho_M=\rho_{M0}\left(\frac{R_0}{R}\right)^3 ,\;\;
\rho_{\phi}=\rho_{\phi
0}\left(\frac{R_0}{R}\right)^{3(1+\omega)} ,
\end{equation}
with the subscript $0$ taken to always mean current value. As to
the varying cosmological term $\Lambda$, we generally assume
$\kappa^2\rho_{\Lambda}=\Lambda=\Lambda_0(R_0/R)^{n}$, where $n$
can, in principle, take on the values 1, 2 and 3. However, only
the value $n=1$ corresponds to a model with additional dynamical
content relative to the constant-$\omega$ usual models, the
cases $n=2$ and $n=3$ just reducing to the Pietro-Demaret model
[11] with the cosmological dimensionless parameter (see below)
$\Omega_k$ replaced for $\Omega_k+\Omega_{\Lambda}$ and
$\Omega_M$ replaced for $\Omega_M+\Omega_{\Lambda}$,
respectively. We then take for the most general conservation law
for $\Lambda$
\begin{equation}
\kappa^2 \rho_{\Lambda}\equiv\Lambda=\Lambda_0 \left(\frac{R_0}{R}\right)
.
\end{equation}
In any event, however, $\Lambda$ can be taken to represent the
energy density of the field $\phi$ corresponding to a particular
value of parameter $\omega$ ($\omega=-2/3$ in the chosen
conservation law (2.8)), so that if we would allow both signs
for $\Lambda$ and $\rho_{\phi}$ then keeping simultaneously
$\Lambda$ and $\rho_{\phi}$ in the field equations would just be
redundant. Nevertheless, one still can consistently consider
field equations with $\Lambda$ and $\rho_{\phi}$ simultaneously
provided that we restrict their values to be either (i)
$\Lambda>0$, $\rho_{\phi}<0$, or (ii) $\Lambda<0$,
$\rho_{\phi}>0$. In what follows, we shall confine ourselves to
just case (i), looking at the quantity
$\rho_v\equiv\rho_{\Lambda}+\rho_{\phi}$ as the total vacuum
energy which will always be taken to be $\rho_v\geq 0$. Bearing
in mind such a restriction, we choose Eqs. (2.7) and (2.8) as
our conservation laws and introduce then the following
dimensionless cosmological parameters [11]
\begin{equation}
\Omega_k=-\frac{k}{R_0^2 H_0^2} ,\;\;
\Omega_{\phi}=\frac{\kappa^2\rho_{\phi 0}}{3H_0^2} ,\;\;
\Omega_M=\frac{\kappa^2\rho_{M0}}{3H_0^2}
\end{equation}
\begin{equation}
\Omega_{\Lambda}=\frac{\Lambda_0}{3H_0^2} ,
\end{equation}
which should satisfy the {\it quadrilateral} constraint
\begin{equation}
\Omega_0=1-\Omega_k=\Omega_M+\Omega_{\phi}+\Omega_{\Lambda} ,
\end{equation}
rather than the constraint implied by the usual cosmological
triangle. From the conservation laws and the first field
equation, we obtain a differential constraint on the scale
factor
\[\dot{R}^2=R_0^2 H_0^2\left[\Omega_M\frac{R_0}{R}\right.\]
\begin{equation}
\left.+\Omega_k+\Omega_{\phi}\left(\frac{R_0}{R}\right)^{1+3\omega}
+\Omega_{\Lambda}\frac{R}{R_0}\right] ,
\end{equation}
and from the relations between $R$ and $V$, and $\dot{R}$ and
$V'$, derived by di Prieto and Demaret [11], the following
generalized constraint on the scalar quintessence potential
\[\left(\frac{V'}{V_0 '}\right)^2=
\Omega_M\left(\frac{V}{V_0}\right)^{\frac{\omega+2}{\omega+1}}\]
\begin{equation}
+\Omega_{\phi}\left(\frac{V}{V_0}\right)^2
+\Omega_k\left(\frac{V}{V_0}\right)^{\frac{3\omega+5}{3(\omega+1)}}
+\Omega_{\Lambda}\left(\frac{V}{V_0}\right)^{\frac{3\omega+4}{3(\omega+1)}}
,
\end{equation}
which has been obtained assuming that $\omega\neq -1$, with
\begin{equation}
V_0=\frac{3}{2}(1-\omega)H_0^2\Omega_{\phi} ,\;\; V_0 '=\pm
H_0\sqrt{\frac{1}{2}\left(1-\omega^2\right)V_0} .
\end{equation}
Constraint (2.13) is of course a generalization from constraint
(1.1) and reduces to this when we set $\Omega_{\Lambda}=0$.

Although again the exponential [12] and cosine [13] potentials
cannot satisfy the constraint (2.13) even if we relax the
condition of nonclosedness implied by nucleosynthesis and
supernova observations, we note that there exist some
inverse-power law potentials, similar to those used in the
literature [12], which satisfy our field equations and
conservation laws. Thus, if we set
$\Omega_{\phi}+\Omega_k=\Omega_{\Lambda}=0$, and $\Omega_M=1$,
we have as a solution to constraint (2.13)
\[V=V_0\left(\frac{\phi_0}{\phi}\right)^4 ,\]
for $\omega=-1/3$. On the other hand, setting
$\Omega_{\phi}+\Omega_{\Lambda}=\Omega_M=0$, we obtain another
solution
\[V=V_0\left(\frac{\phi_0}{\phi}\right)^2 ,\]
for $\omega=-2/3$. Even though they correspond to particular
constant values of $\omega$, these two potentials could still be
implemented in the realm of high energy physics. They have the
same general form as the Ratra-Peebles potential [12], though
this does not actually require that $\omega$ be assumed
constant. Indeed, the above two potentials may be regarded to
belong to a potential family $V=V_0(\phi_0/\phi)^{6(1+\omega)}$
which can be related to the Ratra-Peebles potential by the field
transformation $\phi_{RP}\rightarrow\alpha\phi^{1-3\omega}$,
with $\alpha$ a suitable dimensional constant. In addition,
there are other allowable potentials for the field $\phi$ which
are solution to Eq. (2.13) for particular values of the
quintessence parameter $\omega$, without imposing any
restriction on the cosmological parameters $\Omega_j$. Thus, for
$\omega=-1/3$, we get a family of generalized quintessence
potentials which are given in terms of the Jacobian elliptic
functions, $J_e$ [14],
\begin{equation}
V(\phi)=V_0\left\{J_e\left[\beta_0(\phi-\phi_0)\right],m\right\}^{-4}
,
\end{equation}
where $\beta_0\equiv\beta_0(\Omega_i)$ and $m\equiv
m(\Omega_i)\leq 1$ (with $i$ some elements of the set
$\{\phi,\Lambda,M,k\}$) are dimensionless quantities whose
explicit shape will depend on the particular function $J_e$
being considered. For reasons which will become clear in the
next section, of particular interest for reproducing a suitable
accelerating model of the universe are the elliptic functions
$J_e={\rm cn}$, for which
\[\beta_0=
\frac{1}{2}\sqrt{\frac{\Omega_M}{\Omega_{\phi}}\left(1
+\frac{\Omega_{\Lambda}}{1-\Omega_M}\right)},\;\;
m=\frac{\Omega_{\Lambda}}{1+\Omega_{\Lambda}-\Omega_M} ,\]
$J_e={\rm nc}$, for which
\[\beta_0=
\frac{1}{2}\sqrt{\frac{\Omega_M}{\Omega_{\phi}}\left(-1
+\frac{\Omega_{\Lambda}}{1-\Omega_M}\right)},\;\;
m=\frac{1-\Omega_M}{1+\Omega_{\Lambda}-\Omega_M} ,\] and
$J_e={\rm sd}$, for which
\[\beta_0=
\frac{1}{2}\sqrt{\frac{\Omega_M}{\Omega_{\phi}}} ,\;\;\;
m=\frac{1}{2}\left(\frac{1-\Omega_{\Lambda}}{\Omega_M}\right)
.\] Note, furthermore, that in the limiting case that
$m\rightarrow 1$, the ${\rm cn}$-solution becomes a ${\rm
sech}$-solution for an open universe, the ${\rm nc}$-solution
becomes a $\cosh$-solution for $\Omega_{\Lambda}=0$, whereas the
${\rm sd}$-solution reduces to a $\sinh$-solution for
$\Omega_{\Lambda}=0$ (i.e. the potential first found by Di
Prieto and Demaret [11], as it should be expected.)

On the other hand, for $\omega=-2/3$, we obtain the potentials
satisfying constraint (2.13):
\begin{equation}
V(\phi)= V_0\left[A_{\pm}\sinh(\phi-\phi_0)
+B\cosh(\phi-\phi_0)-C\right]^{-1},
\end{equation}
where
\begin{equation}
A_{\pm}=
\pm\sqrt{\frac{\Omega_M}{\Omega_{\phi}+\Omega_{\Lambda}}}\cosh\delta_0
'\mp\frac{\Omega_k^2 e^{-\sigma_0
'}}{4\Omega_M\left(\Omega_{\phi}+\Omega_{\Lambda}\right)}
\end{equation}
\begin{equation}
B=
\sqrt{\frac{\Omega_M}{\Omega_{\phi}+\Omega_{\Lambda}}}\sinh\delta_0
'+\frac{\Omega_k^2 e^{-\sigma_0
'}}{4\Omega_M\left(\Omega_{\phi}+\Omega_{\Lambda}\right)}
\end{equation}
and
\begin{equation}
C=\frac{1}{2}\frac{\Omega_k}{\Omega_{\phi}+\Omega_{\Lambda}} ,
\end{equation}
with $\delta_0 '$ and $\sigma_0 '$ as given by $\delta_0$ and
$\sigma_0$ in Eq. (1.3), but with $\Omega_{\phi}$ replaced for
$\Omega_{\phi}+\Omega_{\Lambda}$. Clearly, for
$\Omega_{\Lambda}=0$, solutions (2.16) reduce to solutions
(1.3).

Finally, we can also have solutions to (2.13) for any $\omega$,
satisfying $-1<\omega\leq 0$. These solution are in turn
obtained for particular values of the cosmological parameters
$\Omega_j$. Thus, setting $\Omega_k=\Omega_{\Lambda}=0$, we have
\begin{equation}
V= V_0\sinh^{\frac{2(\omega +1)}{\omega}}\left[\pm
\left(\frac{\sqrt{\Omega_M}\omega}{2(\omega
+1)}\right)(\phi-\phi_0)\right] ,
\end{equation}
and for $\Omega_k=\Omega_M=0$,
\begin{equation}
V= V_0\sinh^{\frac{6(\omega +1)}{3\omega
+2}}\left[\pm\sqrt{\frac{3\omega +2}{2}}(\phi-\phi_0)\right].
\end{equation}
Potentials (2.20) and (2.21) and at least some of the potentials
in the family (2.15) can actually be regarded as generalizations
from inverse-power law potentials which hold only as
approximations for small values of $\phi-\phi_0$, at large
values of the redshift (see Sec. III). Let us for example
consider the case $Je=sd$ in the family (2.15). For most of its
cosmological evolution $\phi-\phi_0$ remains very small at large
values of the redshift, so that we can approximate $V\propto
(\phi-\phi_0)^{-4}$, except when the potential approaches
current values. Moreover, though for quintessence the most
interesting models are those where $\omega$ is not constant and,
in particular, the tracker models [15], one can see that at
least some of the good properties of these models may be somehow
shared by the potential considered in this paper. Since at least
some of our potentials can be approximated as inverse-power law
functions of the field containing at least one free parameter,
along their primordial evolution these potentials can be
implemented in the realm of high energy physics [18] and linked
to particle models with dynamical symmetry breaking or
nonperturbative effects [19]. On the other hand, it appears that
such potentials can also help solving the cosmic coincidence
problem [20]. Taking again as an illustrative example the
solution $Je=sd$ in Eq. (2.15) we see (see Eq. (3.14)) that if
one sets the initial conditions inmediately after inflation,
i.e. at a redshift $z\sim 10^{28}$, then $\phi\simeq\phi_0$
initially, and $\phi-\phi_0\sim 1$ only now, so explaining why
the quintessence field begins to dominate now. Tracker models
[15] also seem to improve the fine-tuning problem; we hope this
to be the case with some of our potentials as well, in
particular in those models where the total vacuum energy,
$\Omega{\phi}+\Omega_{\Lambda}$, is zero, or generally for the
reasons discussed in Sec. IV.

At first glance studying the cases $\omega=-1/6, -1/3$ could
seem without any physical motivation, as several authors have
already shown [21] that quintessence models based on such values
should be ruled out. However, our quintessence approach is based
on the idea that the vacuum field is splitted into two parts,
one manifested as a varying cosmological constant with positive
energy density, and the other, the quintessence field, always
having negative energy density. This splitting appears to
enlarge the allowed domain of $\omega$-values which are
physically relevant and reproduce the wanted accelerating
expansion of the universe (see Sec. III). This translates, in
particular, in a generalized expression for the decelaration
parameter (Eq. (3.7)), according to which no value of $\omega$
can be ruled out from the onset.

\section{\bf Predicting cosmological models from quintessence
potentials}
\setcounter{equation}{0}

In this section we are going to check whether the considered
solutions satisfying the constraints on the scalar-field
potentials (1.1) and (2.13) are suitable potentials to fullfil
the quintessence's aim, that is, whether such solutions are able
to predict cosmological models matching observations from
supernova Ia. More precisely, we want to compute the luminosity
distance($D_L$)-redshift($z$) relations from the quintessence
potentials satisfying the constraints. Bringing then that
relation, $D_L(z)$, into the magnitude-redshift relation [2],
\begin{equation}
m_B^{{\rm eff}}=\check{M}_B+5\log[D_L(z)],
\end{equation}
where $\check{M}_B=M_B-5\log H_0+25$ is the Hubble-constant-free
$B$-band absolute magnitude at the maximum of a Ia supernova
whose values have to be suitably corrected [2], we could thus
directly compare the predictions from the cosmological
quintessence potentials with the conveniently corrected observed
magnitude. The luminosity distance $D_L$ depends only on the
theory we are working on, and can be written as [22]
\[D_L H_0=\frac{(1+z)}{\sqrt{|\Omega_k|}}
{\it S}\left\{\sqrt{|\Omega_k|}\right.\times\]
\begin{equation}
\left.\int_0^z
dz'\left[\sum_j\Omega_j(1+z')^{3(1+\alpha_j)}
+\Omega_k(1+z')^2\right]^{-\frac{1}{2}}\right\} ,
\end{equation}
in which ${\it S}\{x\}=\sin x$ for $k=+1$, ${\it S}\{x\}=x$ for
$k=0$ and ${\it S}\{x\}=\sinh x$ for $k=-1$, and the parameter
$\alpha_i$ is defined from the energy density so that
\[\rho_j\propto\left(\frac{R_0}{R}\right)^{3(1+\alpha_j)} ,\]
with the subscript $j$ labelling the distinct nongeometrical
contributions, namely $j=M,\phi$ and $\Lambda$, in such a way
that $\alpha_M=0$, $\alpha_{\phi}=\omega$ and
$\alpha_{\Lambda}=-2/3$. Using the definition of the redshift in
terms of the scale factor $R$, we then attain for the argument
of the squared root in the integrand of Eq. (3.2) in case of the
generalized quintessence model,
\[\Pi\equiv\Omega_M\left(\frac{R_0}{R}\right)^{3}\]
\begin{equation}
+\Omega_{\phi}\left(\frac{R_0}{R}\right)^{3(1+\omega)}
+\Omega_{\Lambda}\left(\frac{R_0}{R}\right)
+\Omega_k\left(\frac{R_0}{R}\right)^2 .
\end{equation}
Inserting now the relation [8] $V/V_0=(R_0/R)^{3(1+\omega)}$, we
obtain for $\Pi$:
\[\Pi\equiv\Omega_M\left(\frac{V}{V_0}\right)^{\frac{1}{\omega+1}}\]
\begin{equation}
+\Omega_{\phi}\left(\frac{V}{V_0}\right)
+\Omega_{\Lambda}\left(\frac{V}{V_0}\right)^{\frac{1}{3(\omega+1)}}
+\Omega_k\left(\frac{V}{V_0}\right)^{\frac{2}{3(\omega+1)}} .
\end{equation}
It is now readily realized that $\Pi^{-1/2}$ is the same as $V_0
' V^{1/2}/V' V_0^{1/2}$ as obtained from the constraint (2.13).
Hence,
\[\int_0^z\frac{dz'}{\sqrt{\Pi}}=\]
\begin{equation}
\frac{V_0 '}{3(\omega+1)V_0^{\frac{1}{2}+\frac{1}{3(\omega+1)}}}
\int_{\phi(0)}^{\phi(z)}\frac{d\phi}{V(\phi)^{\frac{3\omega+1}{6(\omega+1)}}}
,
\end{equation}
and therefore we have the following relation between the
luminosity distance and the quintessence potential \[D_L
H_0=\]
\begin{equation}
\frac{(1+z)}{\sqrt{|\Omega_k|}} {\it S}\left\{\frac{V_0
'
\sqrt{|\Omega_k|}}{3(\omega+1)V_0^{\frac{1}{2}+\frac{1}{3(\omega+1)}}}
\int_{\phi(0)}^{\phi(z)}\frac{d\phi}{V(\phi)^{\frac{3\omega+1}{6(\omega+1)}}}\right\}
.
\end{equation}

On the other hand, the deceleration parameter $q_0$ can also be
expressed in terms of the quintessence parameter $\omega$ as
follows [22]
\[q_0=\frac{1}{2}\sum_j\Omega_j(1+3\alpha_j)=\]
\begin{equation}
\frac{1}{2}\left[\Omega_M+\Omega_{\phi}(1+3\omega)-\Omega_{\Lambda}\right].
\end{equation}
Thus, in order to reproduce the wanted accelerating behaviour of
the universe we must have a suitable combination for the values
of the parameters $\Omega_j$ and $\omega$, such that the
resulting value of $q_0$ be negative. We note that for pure
quintessence models with $\Omega_{\Lambda}=0$, any scalar-field
potentials defined for $\omega>-1/3$ could only give a
topological accelerating behaviour if $\Omega_{\phi}<0$.
Clearly, for $\omega=-1/3$, irrespective of the value of
$\Omega_{\phi}$, we always have $q_0>0$, provided $\Omega_M>0$.
In generalized quintessence models with $\Omega_{\Lambda}>0$,
all the above situations predicting deceleration could still
predict acceleration for sufficiently high positive values of
$\Omega_{\Lambda}$.

Let us consider in what follows the cosmological predictions
from the permissible quintessence potentials dealt with in Secs.
I and II for the case that $\Omega_M>0$. We shall start with the
family of potentials (1.4) for $\omega=-1/6$ and
$\Omega_{\Lambda}=0$, in which case Eqs. (3.5) and (3.6) become
\[D_L H_0=\]
\begin{equation}
\frac{1+z}{\sqrt{|\Omega_k|}} {\it
S}\left\{\frac{\sqrt{14|\Omega_k|}}{5\sqrt{\Omega_{\phi}}}
\int_{\phi(0)}^{\phi(z)}d\phi
J_e\left[\alpha_0(\phi-\phi_0),m\right]\right\} ,
\end{equation}
where we have used the definitions (2.14), and
\begin{equation}
q_0=\frac{1}{2}\left(\Omega_M +\frac{1}{2}\Omega_{\phi}\right) .
\end{equation}
Substituting the different Jacobian elliptic functions in the
luminosity distance expression (3.7) and integrating the
resulting expression, it turns out that the only of such
functions for which we obtain a consistent, real $D_L-z$
relation predicting a nonclosed universe with positive vacuum
energy density which is dynamically accelerating is $J_e={\rm
cn}$. In this case the integral in Eq. (3.8) gives
\[\left.\frac{1}{\alpha_0\sqrt{m}}\arccos\sqrt{1-m+\frac{m}{\sqrt{1+z'}}}\right|_0^z
,\] where
\[\alpha_0=\sqrt{\frac{7\Omega_k}{(1-m)\Omega_{\phi}}}\]
\[m=\frac{\Omega_k}{\Omega_{\phi}+2\Omega_k} .\]
For whichever combinations of values of the cosmological
parameters $\Omega_M$, $\Omega_k$ and $\Omega_{\phi}$ satisfying
the triangular constraint $\Omega_k+\Omega_M+\Omega_{\phi}=1$ we
then obtain a dynamically accelerating universe which, according
to Eq. (3.8), is however topologically decelerating if the
vacuum energy density is positive. On the other hand, although
all possible resulting $5\log D_L H_0-z$ plots give a nearly
straight line between $z\simeq 0.01$ and $z\simeq 0.5$ which
appear to slightly accelerate thereafter, the full $5\log D_L
H_0$-interval that corresponds to the $z$-interval of available
Ia supernova observations ($\simeq (0.01-1)$) is always around
6, quite smaller than the observed value $\bigtriangleup
m_B^{{\rm eff}}\simeq 13$ [2,3]. Thus, the scalar-field
potentials (1.4) cannot conform to the observations on
supernovae Ia.

We consider next the potentials for $\omega=-1/3$ which are
given by the general expression (2.15) for $\Omega_M>0$ and
total vacuum energy density $\Omega_{\phi}+\Omega_{\Lambda}\geq
0$. In these cases, Eqs. (3.5) and (3.6) give
\begin{equation}
D_L H_0=
\frac{1+z}{\sqrt{|\Omega_k|}} {\it
S}\left\{\left.\sqrt{\frac{|\Omega_k|}{\Omega_{\phi}}}\phi(z')\right|_0^z\right\}
\end{equation}
\begin{equation}
q_0=\frac{1}{2}\left(\Omega_M-\Omega_{\Lambda}\right) .
\end{equation}
For the 12 different Jacobian elliptic functions involved in
solutions (2.15) we have derived the expressions of the scalar
field in terms of the redshift, $\phi(z)$, in the form of
elliptic integrals of the first kind [14]. It turns out that
only the elliptic functions ${\rm cn, nc}$ and ${\rm sd}$ can
generate non closed universes which are both topologically and
dynamically accelerating. In the case that $J_e={\rm cn}$, we
have for the luminosity distance
\[D_L H_0=\frac{1+z}{\sqrt{|\Omega_k|}}\times\]
\begin{equation}
{\it
S}\left\{2\sqrt{\frac{|\Omega_k|}{\Omega_M\left(1+\frac{\Omega_k}{1-\Omega_M}\right)}}
\left.F\left[\arcsin\sqrt{\frac{z'}{1+z'}},m\right]\right|_0^z\right\},
\end{equation}
with
\[m=\frac{\Omega_{\Lambda}}{1-\Omega_M+\Omega_{\Lambda}},\]
and for $J_e={\rm nc}$,
\[D_L H_0=\frac{1+z}{\sqrt{|\Omega_k|}}\times\]
\begin{equation}
{\it
S}\left\{2\sqrt{\frac{|\Omega_k|}{\Omega_M\left(1+\frac{\Omega_k}{1-\Omega_M}\right)}}
\left.F\left[\arcsin(i\sqrt{z'}),m\right]\right|_0^z\right\},
\end{equation}
with
\[m=\frac{1-\Omega_M}{1-\Omega_M+\Omega_{\Lambda}} .\]
In expressions (3.11) and (3.12) the symbol $F$ denotes elliptic
integral of the first kind [14]. These expressions give $5\log
D_L H_0-z$ plots for different combinations of cosmological
parameters satisfying the quadrilateral constraint
$\Omega_k+\Omega_M+\Omega_{\phi}+\Omega_{\Lambda}=1$ and
conditions $\Omega_k, \Omega_{\phi}+\Omega_{\Lambda}\geq 0$
which represent accelerating expansion with suitably slight
deviations from straight lines occurring at $z\geq 0.5$ only.
However, again as for the case $\omega=-1/6$, the full
variations of $5\log D_L H_0$ along the $z$-observation interval
$\simeq 6$, are quite smaller than the corresponding value
observed in supernovae.

If we take for $J_e$ the function ${\rm sd}$, then the
$z$-dependence of the scalar field can be expressed in the form
\begin{equation}
\phi(z')=\phi_0+ 2\sqrt{\frac{\Omega_{\phi}}{\Omega_M}}
F\left[\arcsin\frac{1}{\sqrt{1+m+z'}},m\right] ,
\end{equation}
where
\[m=\frac{1}{2}\left(\sqrt{1+\frac{4}{\Omega_M}}-1\right) .\]
Inserting the scalar field (3.13) in Eq. (3.10) we get the
wanted $D_L-z$ relation. In this case, this relation gives plots
which show the required accelerating expansion after $z\geq
0.6$, both for flat and open universes, with a full variation of
$5\log D_L H_0$ along the observed $z$-interval of the order 13,
fitting well with the observations [2,3], along all available
$z$-values.

Finally, for the case $\omega=-2/3$, the relation (3.5) reduces
to
\[D_L H_0=\frac{1+z}{\sqrt{|\Omega_k|}}
{\it
S}\left\{\sqrt{\frac{5|\Omega_k|}{\Omega_{\phi}}}\times\right.\]
\begin{equation}
\left.\int_{\phi(0)}^{\phi(z)}\frac{d\phi}{\sqrt{A_{\pm}\sinh(\phi-\phi_0)
B\cosh(\phi-\phi_0)-C}}\right\},
\end{equation}
with the constants $A_{\pm}$, $B$ and $C$ as defined by
expressions (2.17)-(2.19), and the deceleration parameter given
now by
\begin{equation}
q_0=\frac{1}{2}\left[\Omega_M-\left(\Omega_{\phi}+\Omega_{\Lambda}\right)\right]
.
\end{equation}
In order to obtain cosmological models described by real $D_L-z$
relations the conditions of integration [23] in expression
(3.14) must be such that the cosmological parameters satisfy the
following conditions
\begin{equation}
\Omega_{\phi}+\Omega_{\Lambda}
+\frac{1}{2}\left(\Omega_k\pm\sqrt{2\Omega_k}\right)=0
\end{equation}
\begin{equation}
Q^4-4(4-\Omega_M)\left(\Omega_{\phi}+\Omega_{\Lambda}\right)
=\Omega_k^2
\end{equation}
and, either
\begin{equation}
\Omega_k=-1
\pm\sqrt{1
+\frac{1}{4}\left(\Omega_{\phi}+\Omega_{\Lambda}\right)\left(\Omega_M-16\right)}
,
\end{equation}
or,
\begin{equation}
\Omega_k -2\left(\Omega_{\phi}+\Omega_{\Lambda}\right)=0 .
\end{equation}
The parameter $Q$ in Eq. (3.18) has been introduced to simplify
the equation. It is defined in terms of the $\Omega$'s only, as:
\begin{equation}
Q^2=4\left(\Omega_{\phi}+\Omega_{\Lambda}\right)
+\left[2\left(\Omega_{\phi}+\Omega_{\Lambda}\right)+\Omega_k\right]^2
.
\end{equation}

There are two cosmological models which satisfy all these
conditions. They are: {\it Model I}, a flat universe defined by
the parameters $\Omega_M=1$,
$\Omega_{\phi}+\Omega_{\Lambda}=\Omega_k=0$, and {\it Model II},
an open universe defined by the parameters
$\Omega_M=\Omega_{\phi}+\Omega_{\Lambda}=1/4$, $\Omega_k=1/2$.
In both cases $B=0$ and $A_{\pm}\equiv A$ and $C$ become
indeterminate and real. For Model I we obtain $q_0=+1/2$ and,
from expression (3.15),
\[D_L H_0= \sqrt{\frac{2}{|A|}}(1+z)\sqrt{|A|^2-
\left(|C|^2-\frac{1}{1+z}\right)^2}\times \]
\begin{equation}
\left.F\left[\arcsin\sqrt{\frac{|A|+|C|-\frac{1}{1+z'}}{|A|+|C|}},
\sqrt{\frac{|A|+|C|}{2|A|}}\right]\right|_0^z ,
\end{equation}
which always gives rise to a topologically and dynamically
decelerating universe for any combinations of constants $A$ and
$C$ satisfying $|A|+|C|=1$ and the integration condition
$|A|>|C|>0$.

More interesting is Model II, for which one obtains a
topologically uniform expansion, $q_0=0$, and again the
integration condition $|A|>|C|>0$. Taking e.g. $|A|=2$ and
$|C|=1$, it follows from expression (3.15)
\[D_L H_0=
\sqrt{2}(1+z)\sin\left\{\frac{(z+1)\sqrt{2\left[4-
\left(1+\frac{1}{1+z}\right)^2\right]}}{z+2}\times\right. \]
\begin{equation}
\left.F\left[\arcsin\sqrt{2-\left(1+\frac{1}{1+z}\right)},
\frac{1}{2}\right]\right\} .
\end{equation}
Eq. (3.23) gives rise to a $D_L-z$ plot which, in spite of being
associated with a topologically uniform universe, starts
accelerating after $z\simeq 0.5$ in a way that matches the
behaviour observed in distant supernovas. That plot, on the
other hand, consistently shows a variation $\bigtriangleup(5\log
D_L H_0)\simeq 13$ along the observed $z$-interval, fitting well
with the observations at all available values of the redshift.
Therefore, it could be thought that Model II and the solution in
terms of the Jacobian elliptic function ${\rm sd}$ for
$\omega=-1/3$ dealt with above, correspond to "good"
quintessence potentials.

It appears also relevant to perform a similar computation for
the inverse-power law potentials obtained in Sec. II, which are
of the type already proposed in the literature [12] and that
might be justified from high energy physics. For the first of
these potentials, $V=V_0(\phi_0/\phi)^4$, one obtains
\begin{equation}
D_L H_0= (1+z)\sinh\left[\pm\sqrt{\frac{1}{2\Omega_{\phi}}}
\left(1-\frac{1}{\sqrt{1+z}}\right)\right] ,
\end{equation}
and for $V=V_0(\phi_0/\phi)^2$,
\begin{equation}
D_L H_0= (1+z)\sin\left[\pm\sqrt{\frac{1}{\Omega_{\phi}}}
\ln\left(\frac{1}{\sqrt{1+z}}\right)\right] ;
\end{equation}
these two equations are, of course, not valid for $k=0$. It is
interesting to note that in the two cases, we reproduce a
$D_L-z$ plot which nearly matches the observed results,
producing a distinguishable accelerating pattern starting at an
expected $z\simeq 0.5$, and a variation $\triangle(5\log D_L
H_0)$ between 11 and 12, only slightly smaller than what has
been measured, along the observed $z$-interval. Moreover, our
analytical formulae for the luminosity distance-redshift
relation can also be applied to potentials which are defined for
any value of $\omega$, as those given in Eqs. (2.20) and (2.21),
for sufficiently large values of the redshift. Thus, for
potential (2.20) we obtain
\begin{equation}
D_L H_0\simeq
2(1+z)\sqrt{\frac{1+\omega}{3\Omega_{\phi}\Omega_M}}
\left(1-\frac{1}{\sqrt{1+z}}\right) ,
\end{equation}
at large $z$, and for the potential (2.21),
\begin{equation}
D_L H_0\simeq
(1+z)\sqrt{\frac{2(3\omega+2)}{3(\omega+1)\Omega{\phi}}}
\left(\sqrt{1+z}-1\right) ,
\end{equation}
at large $z$ for $\omega<-2/3$. It can be checked that these
functions give $D_L -z$ plots which show a nearly uniform
expansion at the allowed sufficiently large values of the
redshift.

It is worth noticing that for the limiting expressions from
solutions (2.15) and (2.16), obtained by restricting
$\Omega_{\Lambda}=0$ and $\Omega_{\phi}\geq 0$, we either cannot
even obtain a topologically accelerating universe
($\omega=-1/3$), or have no consistent integration procedure
along the complete range of allowed $z$-values that leads to a
definite real luminosity distance ($\omega=-2/3$). Thus, at
least for the particular potentials considered in this work, if
we want to consistently predict cosmological models compatible
with observations on Ia supernovae at large and moderate
redshifts, it appears that quintessence should be generalized in
a way that allows for a more complicate vacuum structure made up
of (i) a time-dependent, "axionic" [18] (as it is pure imaginary
classically) scalar field, $\phi(t)$, with positive pressure and
negative energy density, and (ii) a time-varying positive
cosmological term, $\Lambda(t)$, whose current value $\Lambda_0$
can be quite smaller than that with which it started the
cosmological evolution, in such a way that the full vacuum
energy density is restricted to be
$\rho_{\phi}+\rho_{\Lambda}\geq 0$.

\section{\bf Summary and discussion}
\setcounter{equation}{0}

In this paper we have considered the problem of the quintessence
potential, restricting ourselves to a constant equation of
state, that is: what are the permissible potentials for a
vacuum, time-dependent scalar field predicting cosmological
models that conform to recent observations and, at the same
time, satisfy the constraints imposed by the field equations and
conservation laws, discussing their physical relevance in the
case that the quintessence field corresponds to a non-tracking
constant equation of state. We have generalized the usual
quintessence model, introducing a positive cosmological varying
term, while restricting the scalar-field energy density to be
definite negative and its pressure definite positive in such a
way that the overall vacuum energy density is necessarily
positive or vanishing. Quintessence potentials that satisfy the
above-alluded constraint, both for the usual models and for
models with a varying cosmological term, have been obtained for
particular values of the constant parameter defining the state
equation for the scalar field in the two kinds of models. None
of these potentials have hitherto been used in quintessence,
except those which are given as an inverse-power law. We also
obtain potentials which are given in terms of either hyperbolic
functions or Jacobian elliptic functions, the latter generally
reducing to the former in the limit when the cosmological term
tends to zero. We have also obtained a general expression
relating the luminosity distance with the quintessence
potential, and this has been integrated for all particular
solutions expressed in terms of the redshift. It turned out that
only some of such solutions with nonzero cosmological term are
able to produce cosmological models that conform to an
acceleratingly expanding universe and agree with recent
observations on Ia supernovae.

The cosmological term $\Lambda$ we have used in our generalized
quintessence model depends on the cosmological time through a
linear dependence on the scale factor $R$. This varying
character of $\Lambda$ could {\it a priori} be an useful
property to help solving the known cosmological constant
problem, though not by itself only. Actually, one could not
justify how an initial vacuum energy density of the order
$M_p^4$ may be lowered down to a value smaller than $10^{-47}$
GeV invoking such a dependence; instead, if the $R$-dependence
of $\Lambda$ would be assumed to be the same along the entire
cosmological evolution, then one would need a conservation law
for $\Lambda$ given by $\Lambda=\Lambda_0(R_0/R)^{\gamma}$, with
$\gamma\geq 123/50$. However, the conservation law chosen in
this paper, $\Lambda=\Lambda_0(R_0/R)$, is assumed to hold only
in the late classical cosmological regime; it could well be that
during primordial expansion $\gamma$ had taken on values larger
than $123/50$. For example, taking $\gamma\simeq 3$ during the
primeval expansion up to $R\geq 10^4$ cm, and $\gamma=1$
thereafter, would solve the cosmological constant problem. The
price one would pay to get such a big reward would just be the
allowance for the dynamical content of the quintessence field to
be that of the conventional models, with $\gamma=3$ and
$\gamma=2$ (see Sec. II) during its early evolution.

The conclusions obtained in this work are not general. They just
refer to the solutions that correspond to the particular values
of the quintessence parameter $\omega=-1/6, -1/3, -2/3$ and -1,
and some special cases for any $\omega$. Possibly there will be
other potentials corresponding to different, intermediate values
of $\omega$ that also reproduce the observed cosmological
expansion within the generalized quintessence model. We do not
believe however this to be the case in the realm of the
conventional quintessence model, though more work is obviously
needed to reach a final verdit on this.

\acknowledgements

\noindent For helpful comments and a careful reading of the manuscript,
the author thanks C. Sig\"uenza. Thanks are also due to M. Moles
for enlightening conversations. This research was supported by
DGICYT under Research Project No. PB97-1218.

\end{document}